# An Improved Adaptive Smo for Speed Estimation of Sensorless Dsfoc Induction Motor Drives and Stability Analysis using Lyapunov Theorem at Low Frequencies

Appalabathula Venkatesh
Research Scholar
Electrical and Electronics Engineering,
National Institute of Engineering,
Mysore, Karnataka, India

*Abstract—* In this paper, An Improved Adaptive Sliding Mode Observer (ASMO) is proposed to a Sensorless DSFOC Induction Motor Drives and their stability is analyzed. ASMO is used to estimate the Rotor Speed, Rotor Resistance, Flux, Stator and Rotor currents and the developed electromagnetic Torques. To improve the robustness and accuracy of an adaptive SMO during very low frequency operation, the sliding mode flux observer(SMFO) uses independent gains as the correction terms. The gains of current and rotor flux SMOs are designed using Lyapunov stability theory to guarantee the stability and fast convergence of the estimated variables. In this paper concentrated on Simulink Blocks and their graphs are analyzed with the help of mathematical approach. Also, comparison of results with the basic conventional controllers are done and the results proved that the proposed ASMO method shows excellent Transient and Steady state speed estimation by the Adaptive Estimators, particularly at low frequencies.

## I. INTRODUCTION

The induction machine was independently contrived by NIKOLA TESLA in 1888Q. A Polyphase (3 Phase) Induction drive systems have been co-ordinated into full-sized vehicles, Conveyors, neighbourhood electric vehicles(EV), Elevators, golf cars, motorcycles, industrial (or utility) vehicles and even in the amusement park attractions. These Polyphase drives are designed to attain maximum distance, with Low Slip and maximum power and efficiency.

With the use of 3-Phase Induction Motor we can produce direct rotational motion of wide speeds ranging from 0-18000 rpm and uniform power which is not possible with a conventional IC engine which has a restriction in speed limits. With the replacement of Induction motor in the place of Internal combustion engines(IC) which results in superior Torque/Weight ratio.

Induction motor works on the principle of electromagnetic induction and is a singly excited machine.

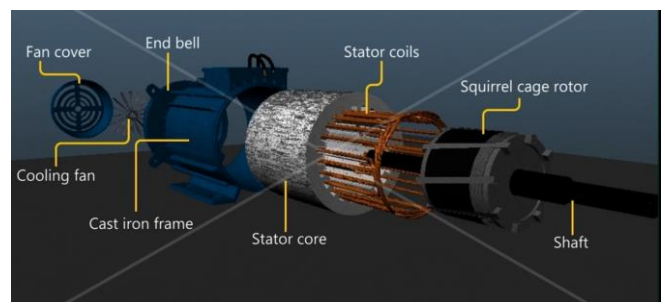

Figure 1.1 Squirrel Cage Induction Motor

It is evident that electrical energy exhaustion of the appliances can be economized by governing the speed of the motors. Fo that reason these three phase variable speed IM drives are uplifted to be used in the industries today as an a fair solution forever for the Decreasing of electricity generation cost.

With the Advancements of Power electronics and Mechatronics technology, a high speed ultrafast switching devices such as IGBTs and Piezeo MEMS were associated with the mechanical systems and more precise motor control strategies, such as Vector or Field control techniques(like Direct or Feedback and Indirect or Feedforward vector Control techniques) , were developed during the last decade. As a result, today IMs can be used in any kind of variable speed applications (called as Asynchronous Motors), even in servomechanism, where high-speed response with respect to the Loads and on the mark of accuracy is required.

Vector or Field control technique is used for the high performance Adjustable drive systems(Variable frequency drives or V/f Control drives). A complex current is mashed-up from the two quadrature components in the vector or field control scheme, one of which is responsible for the flux level in the motor(called as Flux Vector) and the other component, which controls the torque production(called as Torque Vector) in the motor.

The speed and Rotor resistance sensorless estimation concept along with the implementation of Model Reference Adaptive System (MRAS) schemes was studied[1]. It is a well-known fact that the performance of MRAS based speed estimators is betterwhen compared with the other speed estimators with regards to its stability approach and design complexity. Although this thesis is all about ASMO based speed estimators, but it is also the aim of this project is to investigate several speed sensorless estimation strategies for IMs. Conceptual Explanations on the different type of control strategies also were briefly discussed. In the view of simulation works is concerned, the MRAS based speed sensorless estimation schemes chosen in this thesis have been implemented in the Direct Stator Field oriented control (DSFOC) to evaluate the Flux and Speed Observers performance.

Present Research efforts have targeted on replacing the positional encoder on the IM ROTOR shaft and to develop the sensorless drives without effecting their static and dynamic performance. Several speed and dynamic Rotor resistance estimation methods of the sensorless drives have been Introduced. They are classified as machine dynamic-model based methods and high-frequency signal injection methods [4]-[6]. The high-frequency signal injection methods are independent of the machine model. So, these schemes are insensitive to parameters variation and they gives an accurate speed and position estimation particularly at very low and zero stator frequencies. However, they cause high frequency noise which leads to system performance detorioration, in addition to that they require a unusual design which was explained in [6]. Machine model-based schemes gives an accurate and robust speed estimation at extreme high and medium speeds. However, their accuracy and robustness depends mainly on the accuracy of the dynamic IM model for the good operation of the particular drive at very low and zero stator frequencies [4]. For operation at very low stator frequency, dynamic machine model-based methods with more precise models and robust state estimators are to be researched to increase their reliability in a particular region of operations [4]-[5]. Different conventional schemes in the literature survey have been widely used such as MRAS observers [7], adaptive flux observer (AFO) [8]-[12], Extend Kalman filter (EKF) [13], and the sliding mode observers (SMO) [14].

SMO, as a variable structure control system(VSCS), is one of these techniques that gained superior affinity because of its simplicity in design [15], robustness, insensitivity to linear and non-linear parameters variation. MRAS speed-estimation methods which are based on SMO have been introduced in [16] to improve the speed-estimation with superior accuracy when compared with the classical MRAS scheme.

Speed and flux estimation schemes along with ASMO have been discussed in [17]-[22]. These schemes utilizes sliding surfaces which involves stator-current errors in flux estimation. Ingeneral, ASMO schemes use a time-variable full order observer/estimator IM model for flux and current estimation. The estimated speed of these adaptive methods, considered as the final stage of the estimation procedure. Therefore, the errors in the estimated speed due to parameters mismatch and noises reflect directly on the rotor flux estimation, which results in decrease in the accuracy of the flux and speed estimators. These unwanted effects declines the drive performance, specifically at very low and zero stator frequencies, since the fundamental excitation of the motor drive is low. Hence, solution for these issues is helped in concentrating the non adaptive sliding mode observers(Non-ASMO).

SMO(observer) and SMC(controller) for sensorless Direct Torque Controllers(DTC) of IM drives have been discussed in [17-18]. In these research works, the flux estimation was based on Non-ASMO. Specifically in sensorless observers, without speed adaption to provide increased accuracy in a wide speed range operation, which have been presented and compared with others in [19]-[20].

In general, the sensorless IM drive based ASMO are considered as a Adaptive speed observer. In this adaptive speed observer, SMO employs discontinuous correction terms using the current estimation error. Therefore, the errors in the estimated speed reflect specifically on flux estimation and which degrades theaccuracy of state estimators. Unlike the research works in [21]-[22], this paper takes the help of a current model based ASMO for speed and flux estimation to improve their accuracy of estimations, Which can be realized by designing the flux estimation algorithm with a correction term using current estimation error with separate gains are used for current and flux estimation. These gains are designed based on stability conditions of Lyapunov stability theory. This featured solution will improves accuracy of the rotor flux estimators, and which subsequently, Increases the speed and flux estimation accuracy at very low stator frequency operation. The indirect field oriented control (IFOC) for speed control of a sensorless IM drive using the developed estimation algorithms is built by the help of MATLAB or with help of Simulinks.

## II. INSPIRATION FOR FUTURE EXTENSION

Most global optimization problems are nonlinear and thus difficult to solve, and they become even more challenging when uncertainties are present in objective functions and constraints. Efficiency of a Optimization techniques depends upon the search algorithm. Most of the search techniques are single search stage. But in Eagle Strategy there are two searches one is called as Global search and the other is called as the Local Search.

### Think Like a Golden Eagle

"The Victorious strategist only seeks battle after the victory has been won, whereas he who is destined to defeat first fights and afterwards looks for victory."- Sun Tzu

The behaviour of Golden Eagles (*Aquila chrysaetos*) is inspiring. An eagle forages in its own territory by flying freely in a random manner much like then L´evy flights. Once the prey is sighted, the eagle will change its search strategy to an intensive chasing tactics so as to catch the prey as efficiently as possible. There are two important components to an eagle's hunting strategy: random search by L´evy flight (or walk) and intensive chase by locking its aim on the target.

For the Future extension for the Stability analysis of a 3-Phase Induction Motor Rotor Speed and Rotor Resistance we can implement ES-PSO(Eagle Strategy with Particle Swarm Optimization) Technique.

## III. MATHEMATICAL MODELLING

A typical construction of a squirrel cage IM along with the main parts is as shown in Figure 1.1. Its main advantages are the electrical simplicity and mechanical ruggedness, Self starting and the lack of rotating contacts (brushes or Sliprings) requires low maintenance and its capability of better speed regulation.

Before going to analyze any motor or generator it is very much important to obtain the machine in terms of its equivalent mathematical equations. Traditional per phase equivalent circuit has been widely used in steady state analysis and design of induction motor, but it is not appreciated to predict the dynamic performance of the motor. The dynamics consider the instantaneous effects of varying voltage/currents, stator frequency, and torque disturbance. The dynamic model of the induction motor is derived by using a two phase motor in direct and quadrature axes. This approach is desirable because of the conceptual simplicity obtained with two sets of windings, one on the rotor and the other in the stator. The equivalence between the three phase and two phase machine models is derived from simple observation, and this approach is suitable for extending it to model an n-phase machine by means of a two phase machine.

### REFERENCE FRAMES

The required transformation in voltages, currents, or flux linkages is derived in a generalized way. The reference frames are chosen to be arbitrarily such as stationary, rotor and synchronous reference frames. R.H. Park, in the 1920s, proposed a new theory of electrical machine analysis to represent the machine in d – q model. He transformed the stator variables to a synchronously rotating reference frame fixed in the rotor, which is called Park's transformation.[2]-[3]

He showed that all the time varying inductances that occur due to an electric circuit in relative motion and electric circuits with varying magnetic reluctances could be eliminated. In 1930s, H.C Stanley showed that time varying Inductances in the voltage equations of an induction machine due to electric circuits in relative motion can be eliminated by transforming the rotor variables to a stationary reference frame fixed on the stator. Later, G. Kron proposed a transformation of both stator and rotor variables to a synchronously rotating reference that moves with the rotating magnetic field.

### AXES TRANSFORMATION

The per phase equivalent circuit of the induction motor is only valid in steady state condition. It doesn't hold good while dealing with the transient response of the motor. In transient response condition the voltages and currents in three phases are not in balance condition. It is too much difficult to study the machine performance by analyzing the three phases. In order to reduce this complexity the transformation of axes from 3 –Φ to 2 –Φ is needed. Another reason for the transformation is to analyze any machine with 'n' number of phases. Thus, an equivalent per phase model of Induction Motor is accepted and adopted universally, that is called as 'Dynamic d– q model'.

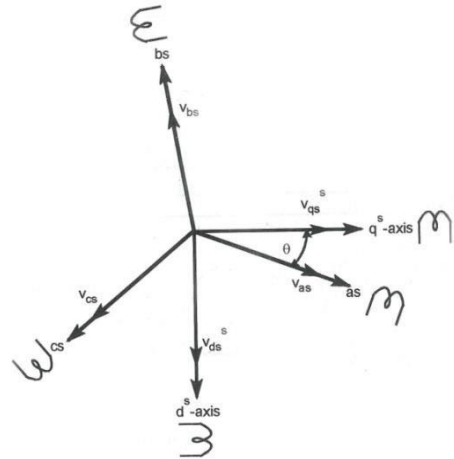

Fig 3.1 3-Φ to 2-Φ Transformation

Consider a symmetrical 3-phase induction machine with stationary $(a_s b_s c_s)$ axis at 120 Degree angle apart. So we needs to transform the 3-phase stationary reference frame $(a_s b_s c_s)$ variables into 2-phase stationary reference frame (d$^s$-q$^s$) variables. Assume that d$^s$- q$^s$ axes are oriented at angle of 90 Degree(Since it is a 2-Phase) which is as shown in Fig 3.1

$$\begin{bmatrix} V_{as} \\ V_{bs} \\ V_{cs} \end{bmatrix} = \begin{bmatrix} \cos\theta & \sin\theta & 1 \\ \cos(120°-\theta) & -\sin(120°-\theta) & 1 \\ \cos(120°+\theta) & \sin(120°+\theta) & 1 \end{bmatrix} \begin{bmatrix} V_{qs}^s \\ V_{ds}^s \\ V_{os}^s \end{bmatrix} \quad (3.1)$$

### DYNAMIC EQUATIONS OF INDUCTION MACHINE

Generally, an IM can be described uniquely in an arbitrary rotating frames i.e.,Stationary reference frame or Synchronously rotating frame. Induction Machine modelling equations with $d^s - q^s$ Axes is given by following dynamic equations which are obtained from the KVL equations to the Dynamic Modelled equivalent circuit of an IM.

$$V_{qs}^s = R_s i_{qs}^s + \frac{d}{dt}\psi_{qs}^s$$

$$V_{ds}^s = R_s i_{ds}^s + \frac{d}{dt}\psi_{ds}^s \quad (3.2)$$

When above equations (3.2) are converted into $(d^e - q^e)$ Axis then above dynamic equations are re-written as follows

$$v_{qs} = R_s i_{qs} + \frac{d}{dt}\psi_{qs} + (\omega_e - \omega_r)\psi_{ds}$$

$$v_{ds} = R_s i_{ds} + \frac{d}{dt}\psi_{ds} - (\omega_e - \omega_r)\psi_{qs} \quad (3.3)$$

For transient studies of a Adjustable/Variable speed drives, it is usually more convenient to simulate an IM when it converts into a stationary reference frame.

Moreover, calculations with stationary reference frame are less complex due to zero sequence frame Component. For small signal stability analysis, a synchronously rotating frame which yields steady-state voltages and currents under balanced conditions.

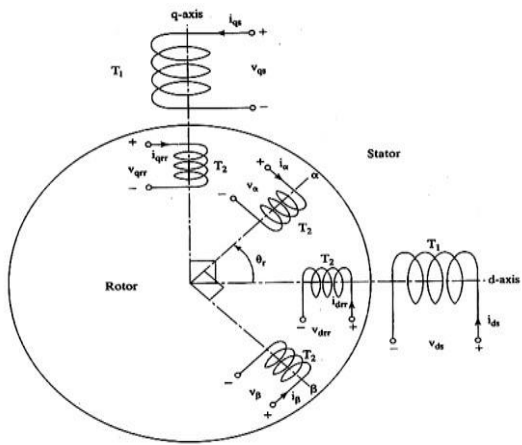

Fig 3.2 Two-phase equivalent diagram of induction motor

Figure 3.3 shows the $q^e$- $d^e$ dynamic model equivalent circuit of induction motor under synchronously rotating reference frame, if $v_{dr} = v_{qr} = 0$ and $\omega_e = 0$ then it becomes stationary reference frame dynamic model.

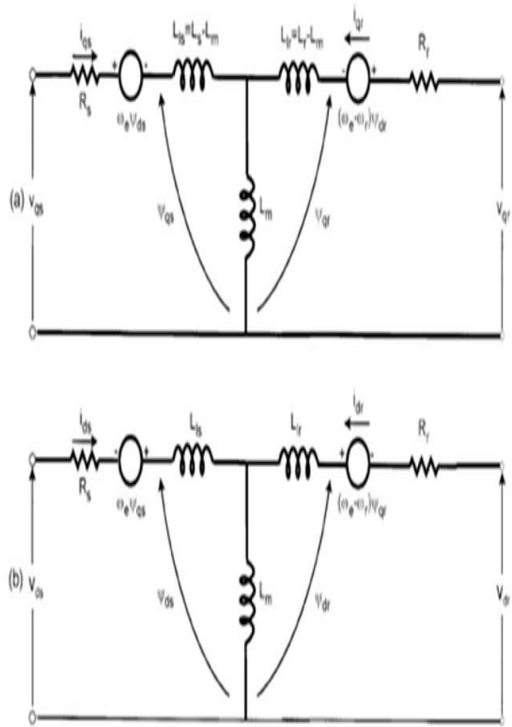

Figure 3.3 Dynamic $d^e$-$q^e$ equivalent circuits of machine

The IM described in stationary reference frame interms of stator currents and Rotor Fluxes are as follows which was described in Reference[6] is designed as follows in the Figure 3.4

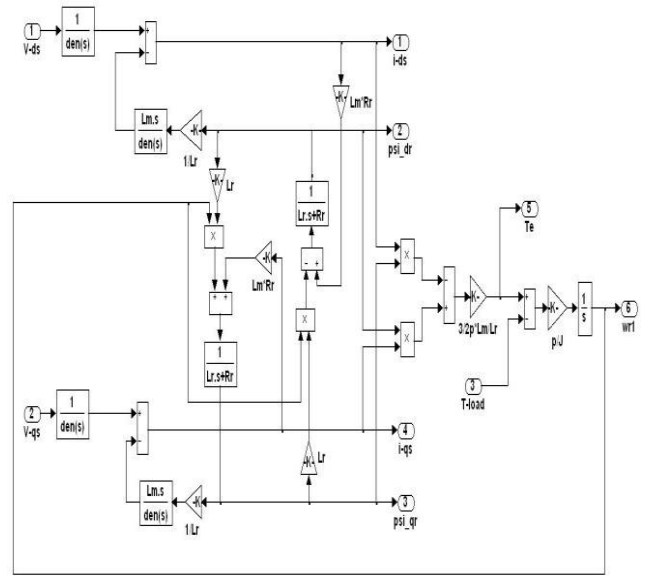

Figure 3.4 Induction Motor Simulation Diagram

| Symbol | Parameter | Value |
|---|---|---|
| $d^e - q^e$ | Synchronously Rotating Reference Frame Direct and Quadrature Axes | |
| $d^s - q^s$ | Stationary Reference Frame Direct and Quadrature Axes (Also known as $\alpha - \beta$ Axes) | |
| F | Frequency | [Hz] |
| $I_{dr}^s$ & $I_{ds}^s$ | $d^s$ Axis Rotor and Stator Currents | [Amp] |
| $I_{qr}^e$ & $I_{qs}^e$ | $q^e$ Axis Rotor and Stator Currents | [Amp] |
| $\theta_e$ | Angle of Synchronously Rotating Frame | [Degree] |
| $\theta_e$ | Rotor Angle | [Degree] |
| $\theta_{sl}$ | Slip Angle | [Degree] |
| $L_r$ & $L_s$ | Rotor & Stator Inductance | [Henry] |
| $L_{lr}$ & $L_{ls}$ | Rotor & Stator Leakage Inductance | [Henry] |
| $L_{dm}$ & $L_{dm}$ | $d^e$ & $q^e$ Axis Magnetizing Inductance | [Henry] |
| $R_s$ & $R_r$ | Stator and Rotor Resistance | [$\Omega$] |
| $V_{dr}^s$ & $V_{ds}^s$ | $d^s$ Axis Rotor and Stator Voltages | [Volt] |
| $V_{qr}$ & $V_{qs}$ | $q^e$ Axis Rotor and Stator Voltages | [Volt] |
| $\psi_{dr}^s$ & $\psi_{ds}^s$ | $d^s$ Axis Rotor and Stator Flux linkage | |
| $\psi_{qr}$ & $\psi_{qs}$ | $q^e$ Axis Rotor and Stator Flux linkages | |

## IV. ADAPTIVE SLIDING MODE OBSERVER (ASMO):

The ability to generate a sliding motion on the error between the measured plant output and the output of the observer ensures that a sliding mode observer produces a set of states. Estimates that are precisely comparable with the actual output of the plant. Standard Testing function (Plant) along with Adaptive SMO is as shown in the below figure 4.1

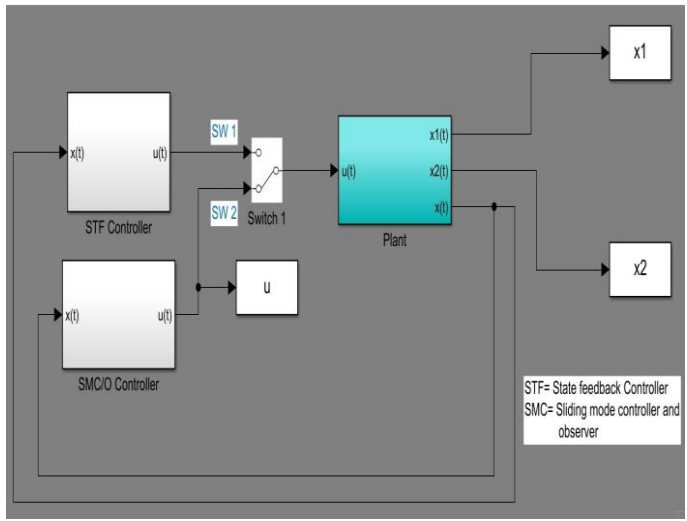

Fig. 4.1 Block diagram of ASMO

### ASMO TESTING FUNCTION:

This simulation is use to demonstrate the robustness property of sliding mode control. Here, second order stable system is consider, which means that states of the system will reach the equilibrium in infinite time.

The transfer function of system is given

$$G(s) = \frac{1}{s^2 + 5s + 6}$$

1. First, we will observe the response for system without disturbance using state feedback controller. In order to do so keep the switch 2 in SW 1 position (open plant). Observer the response, you will find that both the state will reach to the zero in infinite time,
2. In next step we will introduce the sinusoidal disturbance by moving switch from SW1 position to SW 2 position and observer the response of the system; you will find that both the state will oscillate.
3. In the last step will apply the SMC controller to the plant with disturbance and observer the response.

The simulation state responses are as shown below

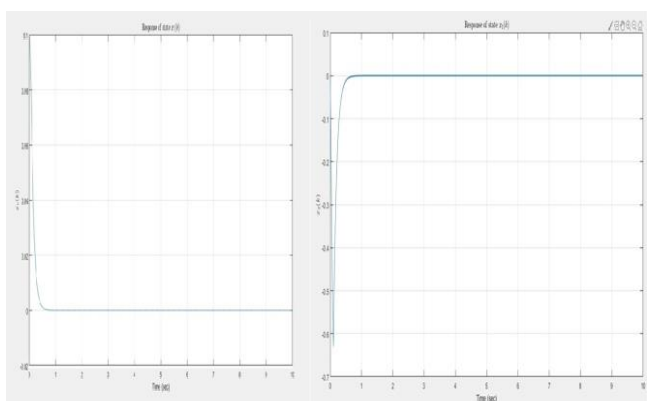

Fig. 4.2 Simulation Results of ASMO Testing Function

## V. PROBLEM FORMULATION

This paper defines the Objective Function as reducing the error between the original Induction Machine Angular Rotor Speed and Rotor Resistance and Rotor Flux and Electromagnetic Torque Developed by the Machine along with respect to standard system to make IM as a stabilized system. So, the objective function can be formulated as follows

$$f(e_x, e_\phi) = \frac{1}{2} e_x^T P e_x + \frac{d}{2} \Delta \omega_r^2 + a^T \alpha a + b^T \beta b \quad (5.1)$$

Where P and Q are positive definite symmetrical matrices. a and b are the vectors which contains the non-zero elements of A and B matrices of Induction Motor modelling equations. α and β are the diagonal matrices with positive elements which determines the speed of adaption of the Optimizing Technique.

Where $\quad e_x = \begin{bmatrix} e_{id}^T & e_{iq}^T & e_{\psi_d}^T & e_{\psi_q}^T \end{bmatrix}$

The above objective function is to be tuned by Model Reference Adaptive System(MRAS) technique. The corresponding Block diagram of MRAS is implemented first. Later, ASMO is applied to optimize the above objective function in Eqn 5.1.

### CANONICAL FORM FOR THE NOMINAL SYSTEM:

Canonical Form of representation of a standard Continuous-Time state space system is as follows

$$\dot{X}_1(t) = A_{11} x_1(t) + A_{12} y(t) + B_1 u(t)$$
$$\dot{Y}(t) = A_{21} x_1(t) + A_{22} y(t) + B_2 u(t) \quad (5.2)$$

Similarly, Adaptive Sliding Mode Observer (ASMO) is represented as follows in State space form is as follows

$$\dot{\hat{X}}_1(t) = A_{11} x_1(t) + A_{12} y(t) + B_1 u(t) + Lv \quad (5.3)$$
$$\dot{\hat{Y}}(t) = A_{21} x_1(t) + A_{22} y(t) + B_2 u(t) - v$$

Where $(\hat{x}_1, y)$ represent the state estimates, $L \in R^{(n-p) \times p}$ is a gain matrix and $v_i = M \, \text{sgn}(\hat{y}_i - y_i)$ where $M \in R$

Error of the system along with ASMO is

$$\dot{e}_x = A_{11} e_x(t) + A_{12} e_y(t) + Lv$$
$$\dot{e}_y = A_{21} e_x(t) + A_{22} e_y(t) - v \quad (5.4)$$

A Lyapunov stability function can be examined to define the stability of the Induction Machine along with the state estimators is represented as follows,

$$V = \frac{1}{2}e_x^T P e_x + \frac{d}{2}\Delta\omega_r^2 + a^T \alpha a + b^T \beta b \qquad (5.5)$$

The Time derivate of above function to guarantee the stability of the system will be $\dot{V} < 0$. Controller gains are adjusted in such a way that $\dot{V}$ is a Negative Definite function. The main aim of ASMO controller parameters adjustments are done to ensure that the state trajectory of the switching function should lies within the specified switching surface. The following state trajectories are used to define the stability of the system.

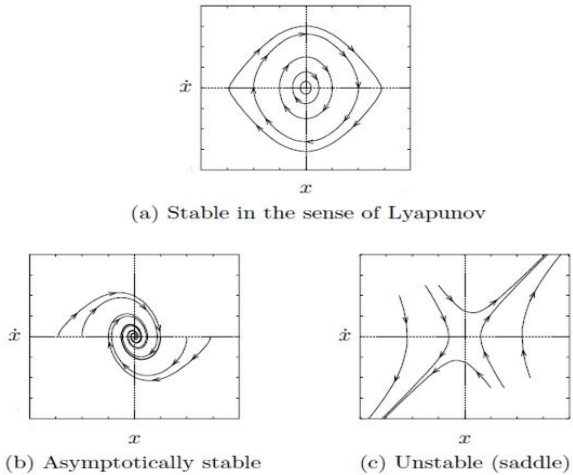

Fig 5.1. State Trajectories to identify the type of stability

Block Diagram representation of Induction Motor along with the State Estimators( Flux and Speed Estimators) along with PID tuners is as represented as follows in figure 5.2

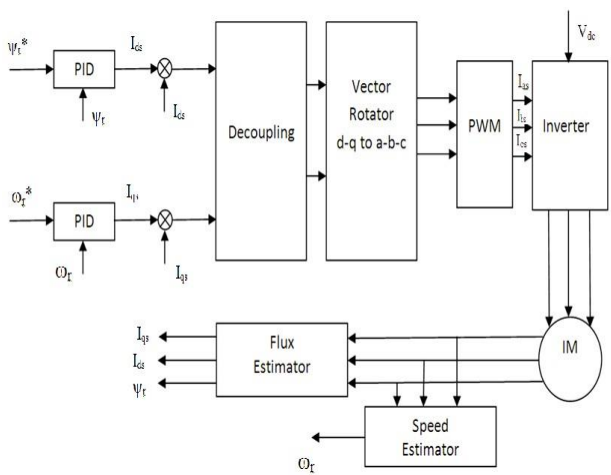

Fig 5.2 Block Diagram of Induction Motor along with Estimators

Simulation Diagram of DSFOC Induction Motor without ASMO is as shown in the below figure 5.3

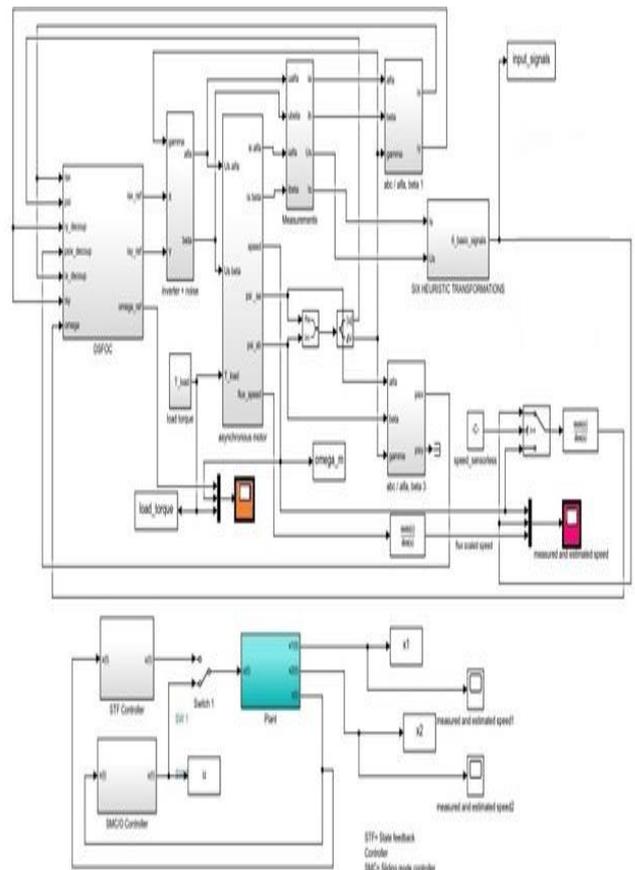

Fig 5.3 Simulation Diagram of DSFOC Induction Motor without ASMO

## VI. SIMULATION RESULTS

The parameter values are considered for the dynamic Induction Machine modelling are as follows:

$L_s = 0.1004$ [Henry], $L_r = 0.0969$ [Henry],

$L_m = 0.0915$ [Henry], $R_r = 1.294$ [Ohm],

$R_s = 1.54$ [Ohm],

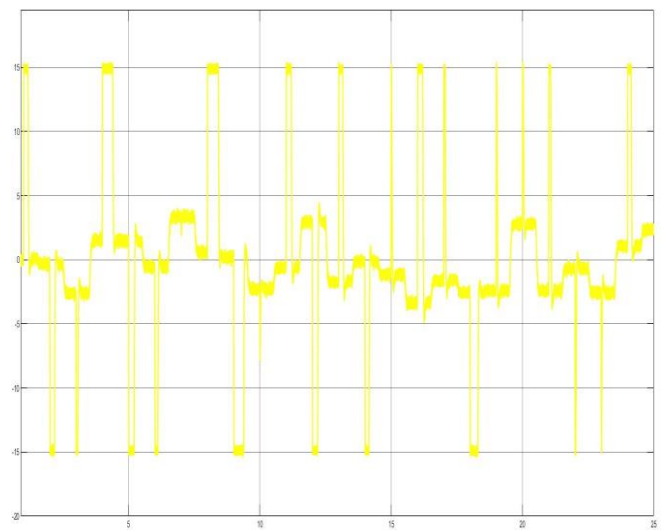

Fig 6.1: Simulation Results of Direct Axis current $I_d$
DSFOC Induction Motor

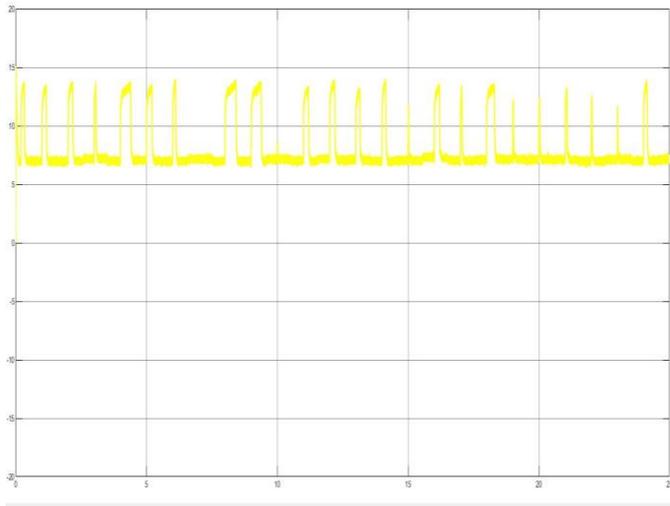

Fig 6.2: Simulation Results of Quadrature Axis current $I_q$ DSFOC Induction Motor

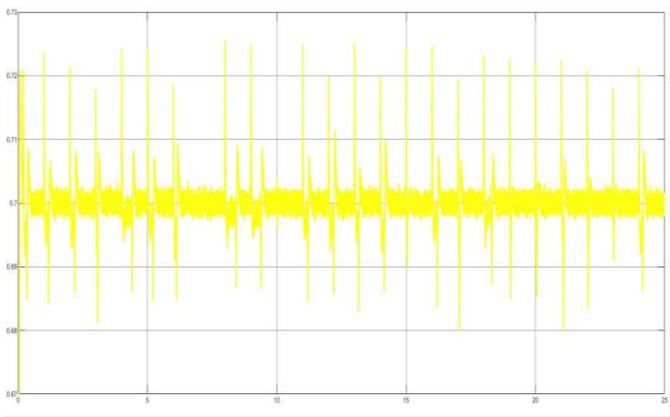

Figure 6.3: Simulation Results of Stator Flux of a DSFOC Induction Motor

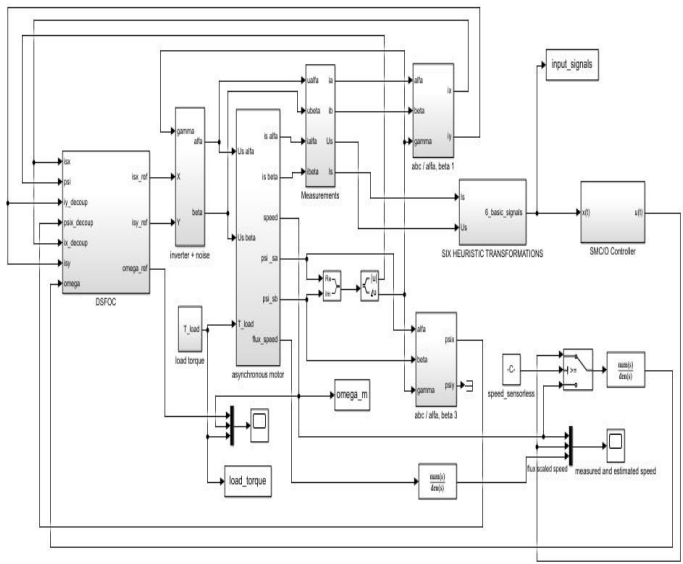

Fig 6.4 Simulation Diagram of DSFOC Induction Motor along with ASMO/C

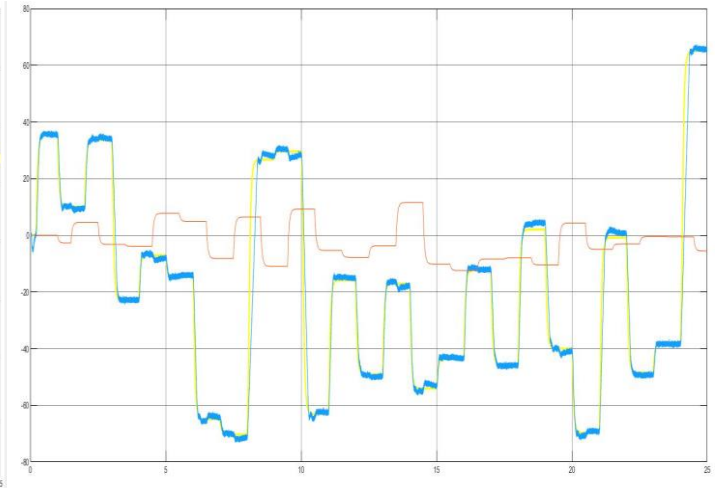

Fig 6.4 Simulation Results of DSFOC Actual Speed, Measured Speed and

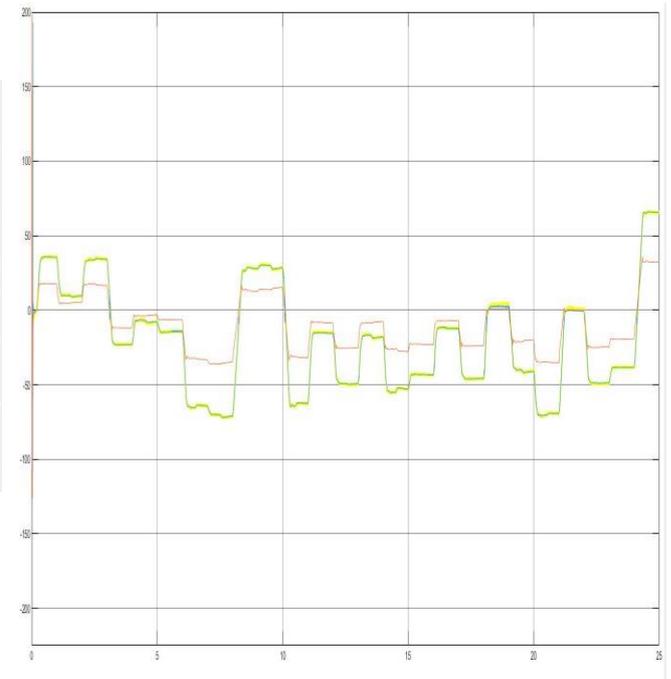

Electromagnetic Torque When Load Torque $T_L = 0$

Fig 6.5 Simulation Results of DSFOC Induction Motor with ASMO Actual Speed, Measured Speed and Electromagnetic Torque When Load Torque

$$T_L = 0$$

In this paper, a new ASMO for Rotor speed and flux and rotor resistance estimation of the sensorless speed DSFOC IM has been designed. Lyapunov stability theorem[25] is utilized for the stator current and Rotor flux estimated to determine the tunable observer gains of the current and flux estimators to guarantee the occurance of the stability of the IM within the sliding surfaces.In dynamic state, the estimator speed are calculated and During the steady state, the state estimators(Flux and Speed) error dynamics behaves as a Reduced order state dynamics and subsequently it is controlled only by the error of the Rotor fluxes. So, the Lyapunov function is selected to estimate the speed of the Rotor and its position. Simulation and Experimental results confirms that the usefulness of the new proposed ASMO for estimating the Rotor Speed of IM at very low and zero stator frequencies. It has been observed that the sensorless drive with the proposed adaptive SMO provides a good performance in comparison with the previous conventional works. But, The SMO algorithm is designed using the dynamic IM mathematical model, Its observability is usually at zero stator frequency. Observability of the IM can be improved by additional stator voltage injection methods and the advanced optimizing techniques like ESPSO[23]-[24] which could be implemented to Increase the effectiveness in reaching the sliding surfaces to attain the better stability which can be considered to be the future extension work.